%% file: hodas12.tex
\documentclass[conference]{IEEEtran}

\usepackage[cmex10]{amsmath}
\usepackage{amssymb}
\usepackage{bbm}
\usepackage{graphicx}
\usepackage{subfigure}
\usepackage{epstopdf}
\usepackage{balance}
\usepackage{hyperref}
\newcommand{\remove}[1]{}
\usepackage{color}

\begin{document}
\title{How Visibility  and Divided  Attention Constrain Social Contagion}
\author{
\IEEEauthorblockN{Nathan Oken Hodas}
\IEEEauthorblockA{Information Sciences Institute\\
4676 Admiralty Way\\
Marina Del Rey, CA 90292\\
nhodas@isi.edu}
\and
\IEEEauthorblockN{Kristina Lerman}
\IEEEauthorblockA{Information Sciences Institute\\
4676 Admiralty Way\\
Marina Del Rey, CA 90292\\
lerman@isi.edu}
}

\maketitle
\begin{abstract}
How far and how fast does information spread in social media? Researchers have recently examined a number of factors that affect information diffusion in online social networks, including: the novelty of information, users' activity levels, who they pay attention to, and how they respond to friends' recommendations. Using URLs as markers of information, we carry out a detailed study of retweeting, the primary mechanism by which information spreads on the Twitter follower graph.
Our empirical study  examines how users respond to an incoming stimulus, i.e., a tweet (message) from a friend, and reveals  that 
dynamically decaying visibility, which is  the increasing cognitive effort required for discovering and acting upon a tweet, \remove{visual saliency} combined with limited attention play dominant roles in retweeting behavior. Specifically, we observe that users retweet information when it is most visible, such as when it near the top of their Twitter feed.
Moreover, our measurements quantify how  a user's limited attention is divided among incoming tweets,  providing novel evidence that highly connected individuals are less likely to propagate an arbitrary tweet.
Our study indicates that the finite ability to process incoming information constrains social contagion, and we conclude that rapid decay of visibility is the primary barrier to information propagation online.
\end{abstract}


\input{introduction-final} 
\input{empirical-final}
\input{related-work-final}
\input{discussion-final}

\section*{Acknowledgment}
This material is based upon work supported by the National Science Foundation under Grant No. IIS-0968370,  the Air Force Office of Scientific Research under Contract No. FA9550-10-1-0102 and by the Air Force Research Laboratory (AFRL) under Contract Number FA8750-11-C-0127.



\end{document}

%% file: introduction-final.tex
\section{Introduction}

Psychologists have identified attention as the mechanism that controls our ability to process incoming stimuli and make decisions about what activities to engage in~\cite{Kahneman73,Rensink:1997vj,Pashler:1998ug}.
Computer science researchers have recently recognized the importance of attention in explaining human behavior online~\cite{Goldhaber97,Weng:2012dd}.
Attentive acts, such as reading a tweet, browsing a web page, or responding to email, require mental effort, and the human brain's capacity for mental effort is limited due it its finite energy resources~\cite{Muraven:1998ta,Baumeister:2008ge,blus:2008tn,Sarter:2006dl,Smit:2004jo}.
As a consequence, the more stimuli people \remove{receive} have to process, the smaller the probability they will respond to any one stimulus, since they must divide their finite attention over all incoming stimuli. In social media, the phenomenon of \emph{divided attention} was shown to limit the number of meaningful interactions people can maintain online~\cite{Goncalves11} and govern  what~\cite{Counts11, blus:2008tn} or who~\cite{Huberman-attention,Backstrom:2011ts} people allocate their effort to.
We show that divided attention also plays a critical role in social contagion,  the process by which information spreads online from person to person through follower relations. Our study quantifies the degree to which a user's attention is divided and the effect this has on the retweeting behavior.

To conserve mental resources (and avoid mental exhaustion from routine tasks) humans have developed a number of strategies, which can be expressed as a \emph{principle of least effort}~\cite{Kool:2010he,Kurniawan:2011bi,Tavares:2012co}: organisms will adapt their behavior to utilize a minimal amount of energy, with the constraint of achieving minimally satisfactory results.  In other words, tasks  requiring small amounts of time or effort are more likely to be chosen than tasks requiring large amounts of time or effort but providing minimal additional reward.  This is also known as `satisficing'~\cite{Agosto:2001dt}.  In the social media domain this implies that memes that require little effort  to be discovered by users are more likely to be propagated than memes requiring more effort to be discovered.
We link the effort required to  perceive something to its `\emph{visibility}.' High `visibility' memes take little time and effort to be discovered, while low `visibility' memes require more time and energy to be perceived. Moreover, as the visibility of a meme decays, it is less likely to be discovered and propagated.  We take `visibility' to be a meme-specific generalization of visual saliency and top-down perception processes for discovering of an item within the visual field~\cite{Parkhurst:2002vo,Itti:2001wa,Melcher:2011eq,Fine:2009jk}. We quantify this effect empirically through a statistical study of retweeting.

\remove{
 Just as individual attention is finite, so is collective attention.
Competition for finite collective attention was shown to be an important factor in the production~\cite{Wu10attention} and evolution of popularity~\cite{Wu07,Ratkiewicz10} of online content.
}

We study the spread of information on the Twitter follower graph, using URLs embedded in tweets as distinct markers of information.  Users could be exposed to information from sites external to Twitter, such as the New York Times, and decide to broadcast this information to their followers by tweeting a URL linking to the information.
Alternately, users could receive the  information via Twitter 
and rebroadcast it to their  followers by retweeting.
Our study of human response dynamics~\cite{Eckmann:2004wm,citeulike:5280499} examines how Twitter users respond to incoming stimuli and shows that retweeting behavior can be properly understood as an interplay between the dynamic visibility of the stimuli and the divided attention of users.
Twitter organizes a user's queue, i.e., tweets from friends the user sees upon visiting Twitter, chronologically, with the most recent tweets at the top of the queue.
Because users' attention is limited, they inspect only a finite portion of the queue, usually starting at the top~\cite{Huberman:1998eq}.
This gives tweets residing at the top of the queue  the highest visibility, but visibility decays as new tweets arrive, pushing older ones farther down the queue.

We observe the universal pattern that the deeper a tweet is in a user's queue, the less likely the user is to discover the tweet. Our empirical analysis quantifies how people tend to retweet information when it is easiest to discover, which is immediately after a friend has tweeted it.
The visibility of information quickly decays as friends add new tweets to a user's queue, and the rate of decay
becomes more rapid as the number of friends a user has grows, making individuals with a high in-degree (many friends) less likely to propagate an arbitrary tweet than low in-degree individuals.
We propose that  decay of visibility combined with divided attention  explains why most information cascades in social media fail to reach epidemic proportions~\cite{Versteeg11}.


This paper makes the following contributions.
We carry out (Section~\ref{sec:frf}) statistical analysis of URL retweeting activity by splitting users into populations based on relevant features, such as attention, and performing statistical analysis of user behavior of within each population.
 We show (Section~\ref{sec:attention}) that the probability of responding to any friend's tweet is also inversely proportional to the number of friends, due to attention being statistically divided across all incoming tweets.
Moreover, we show (Section~\ref{sec:visibility}) that a user's response to a single exposure to a URL by a friend decays in time, with median response time inversely proportional to the number of friends she follows.
We interpret these findings  as being explained from the perspective of visibility and divided attention,  rather than ``novelty decay''~\cite{Wu07}.

%% file: empirical-final.tex
\section{Empirical study}
\label{sec:empirical}
Twitter ({\tt www.twitter.com}) is a popular social networking site that allows registered users to post and read short text messages, called tweets, and follow the activity of other users.  The Twitter interface employs a first-in-first-out queue, so the most recently tweeted content is at the top of the screen, with older information ordered chronologically.  A user can also retweet the content of another user's post, analogous to forwarding an email. A tweet often contains a URL to  external web pages.  We use these URLs as a distinct markers of information. When a user tweets a URL, this information broadcasts to all of her followers, who may subsequently retweet it, broadcasting the information to their followers, and so on. Thus URLs \emph{cascade} through the follower graph.  Although a number of independent twitter clients exist, we do not distinguish them, because most of them utilize a the first-in-first out queue.
We use data collected from Twitter to study mechanics of social contagion which give rise to information cascades.


\subsection{Data collection}
Twitter's Gardenhose API provides access to a portion of real time user activity -- roughly 20\%-30\% of all user activity at the time data was collected.
We used this API to collect tweets over a period of three weeks in the Fall of 2010.
We retained tweets that included a URL in the body of the message, usually shortened by some URL shortening service, such as bit.ly or tinyurl.
In order to ensure that we had the complete tweeting history of each URL, we used Twitter's search API to retrieve all tweets containing  that URL.
For each user who sent a tweet containing a URL, we used the REST API to collect friend and follower information for that user.
This data collection process resulted in more than 3 million tweets which mentioned 70K distinct shortened URLs.
 There were 816K users in our data sample, but we were only able to retrieve follower information for some of them, resulting in a follower graph with almost 700K nodes and over 36 million edges.

Retweeting activity in our sample encompassed diverse behaviors from organic spread  of newsworthy content to orchestrated human and bot-driven campaigns of advertising and spam.
 We recently demonstrated a method to automatically classify these behaviors~\cite{Ghosh2011snakdd}, using   two information theoretic features computed from characteristics of the retweeting dynamics, user entropy and time entropy.
 The first feature is the entropy of the distinct user distribution, and the second feature is the entropy of the distinct time interval distribution.
 We showed that these two features alone were able to accurately separate activity into meaningful classes.
 High user entropy implies that many different people retweeted the URL, with most people retweeting it once.
 High time interval entropy implies the presence of many different time scales, which is a characteristic of human activity.
 In this paper, we focus  on those URLs from the data set which are characterized by high ($>3$) time interval entropy and user entropies at least as large as the user entropies.
  These parameter values are associated with organic spread of newsworthy content by many individuals.
This `spam-filtered' data set contained 2072 distinct URL's retweeted a total of 1337K \remove{1,336,981} times by 487K \remove{486,906} distinct Twitter users.


\subsection{Tweet Response Curves}~\label{sec:frf}
We use time stamps contained in tweets combined with the follower graph to track when users are exposed to URLs and when they retweet them.
In the present context, we define a retweet to be anytime a user tweets a URL that had previously appeared in her Twitter queue.
We consider user $u$ to be exposed to a URL if at least one of the people $u$ follows sends at least one  tweet containing the URL. Using the follower graph, we construct a set of `friends,' denoted $F^R_u$, who the user $u$ follows, and a set of `followers,' $F^O_u$, who follow $u$.  The \emph{friend response function} measures how users behave when multiple friends expose them to the same URL. The \emph{exposure response function} measure how users behave upon receiving multiple tweets containing the same URL, even if sent by the same user.  These two functions provide a picture of how user attention is divided among friends and incoming tweets.

\subsubsection{Friend response function}
We measure the friend response function for retweeting a URL by parsing the time-series of tweets as follows. We chronologically order all $N_j$  tweets  containing URL$_j$, forming an ordered list $C_j=\oplus_i^{N_j}\{t_i,u_i\}$, where $t_i$ is the time of the $i^{th}$ tweet, which is made by user $u_i$.  If multiple tweets occur simultaneously or a user tweets the same URL multiple times, time-stamps and users may appear in $C_j$ multiple times.  We use the notation $C_j[u]$ to indicate the set of tweets containing URL$_j$ made by user $u$. For each user in the time-series, after determining the first time they tweet or retweet URL$_j$, i.e., $t^*_{u,j} =\inf_t C_j[u]$,  we establish how many users in their friend set, $F^R_u$ also sent the URL before time $t^*_{u,j}$, giving the set of prior tweets of URL$_j$ exposed to $u$:
\begin{displaymath}
V^+(u,j) = F^R_u\bigcap_{i:t_i<t^*_{u,j}}\{u | u \in C_{j}\},
\end{displaymath}
where $\bigcap$ denotes the intersection of the users who previously tweeted as described.
Some users are exposed to a URL, but they do not subsequently retweet it.  This set of  `watchers' is constructed by taking the set of all users who were exposed to the URL and removing the set of users who tweeted the URL,
\begin{displaymath}
\mathcal{W}_j=\left\{u^\prime\in \bigcap_{u:u \in C_j} F^O_u \bigg / u^\prime \notin C_j\right\},
\end{displaymath}
where $\bigg /$ denotes the set difference operator.

The set of friends who exposed user $u$ to URL$_j$, given that $u$ did not tweet URL$_j$, is then
\begin{displaymath}
V^-(u,j) = F^R_u \bigcap C_j, \text{ given } u\in \mathcal{W}_j.
\end{displaymath}
Notice  no restriction on time exists in $V^{-}$.  We then can calculate the empirical probability that a user will retweet a URL$_j$ given that $n$ friends tweeted it previously, which is the friend response function:
\begin{align}\label{eq:friendresponsefunction}
P_f(n) &= \\ \nonumber
 &\frac{\sum_{u,j}\mathbbm{1}_{|V^+(u,j)| = n}}{\sum_{u,j}\mathbbm{1}_{|V^+(u,j)| = n}+\mathbbm{1}_{u\in\mathcal{W}}\mathbbm{1}_{|V^-(u,j)| = n}},
\end{align}
\noindent
where the characteristic function  $\mathbbm{1}_{\chi} = 1$ if the condition $\chi$ is true and 0 otherwise, and $|V(u,j)|$ is the cardinality of set $V(u,j)$.  The retweet probability  for users with specific characteristics, such as having many or few friends, is calculated by further restricting the users counted in \eqref{eq:friendresponsefunction}, expressed as
 \begin{align}\label{eq:friendresponsefunctionGeneral}
P_f(n|\chi) &= \\\nonumber
&\sum_{u,j}\mathbbm{1}_{u\in\chi}\mathbbm{1}_{|V^+(u,j)| = n}\bigg/ \\
\sum_{u,j}\mathbbm{1}_{u\in\chi}&\mathbbm{1}_{|V^+(u,j|\chi)| = n}+\mathbbm{1}_{u\in\chi}\mathbbm{1}_{u\in\mathcal{W}}\mathbbm{1}_{|V^-(u,j|\chi)| = n},
\end{align}
where $\chi$ is the restricted feature of the subpopulation of users of interest, and $\mathbbm{1}_{u\in\chi}$ is the characteristic function, which is 1 if user $u$ has feature $\chi$ and 0 otherwise.

\begin{figure}[htbph] 
   \centering
   \subfigure[Friend response function, averaged over all users]{
  	 \includegraphics[width=3.5in]{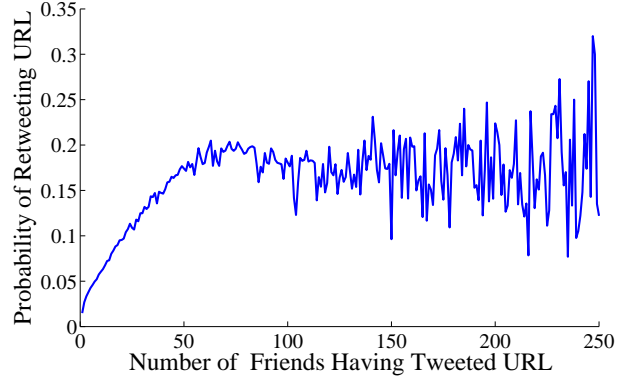}
 	  \label{fig:twitterFriendResponseAll}
   }
    \subfigure[Friend response functions, grouped by number of users' friends]{
  	 \includegraphics[width=3.5in]{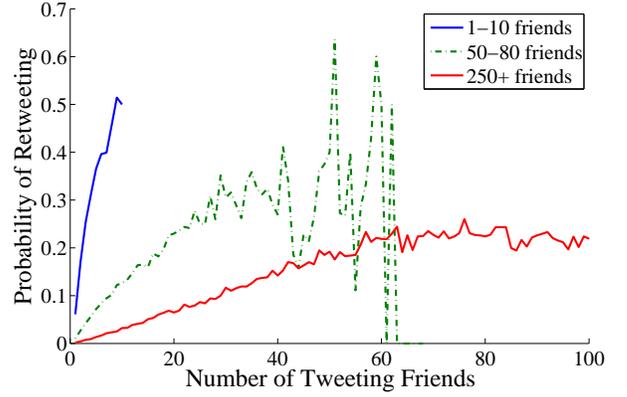}
 	  \label{fig:twitterFriendResponseBreakdown}
   }
   \caption{The \emph{friend response function} is the probability of retweeting as a function of the number of friends who have previously tweeted the URL. (a) The probability of retweeting a URL increases as more friends tweet the URL.  (b) By breaking down users into classes based on the number of  friends they have, and measuring the retweet probability separately for each sub-population, we see that having more friends reduces the user's sensitivity to incoming tweets.  Although the friend response functions may appear to subtly decrease above  $\sim50$ friends, this is an artifact of  finite-time data collection and inhomogeneities in the underlying  populations as a function of received retweets, {\emph{not}} a true decrease in likelihood of retweeting at high exposure.
  }
   \label{fig:friendresponse}
 \end{figure}

Figure~\ref{fig:twitterFriendResponseAll} shows the friend response function that has been averaged over all users.  This function is similar in form to that observed in previous works~\cite{Romero:2011um,Versteeg11}. For example, Romero et al.~\cite{Romero:2011um} measured the probability that Twitter users will adopt a hashtag given $n$ of their friends had previously used it.  They found that the probability of hashtag adoption increased with the number of adopting friends, up to a point, and then strongly declined. Unlike the hashtag-based friend response function, which peaked after about five friends have adopted it, the URL-based friend response function peaks after about 50 friends have tweeted a URL.
Preliminary simulation work shows that the drop-off in the friend response function is best explained by the finite-time of the data collection period and differences between the population of users who receive only a few retweets and those who can receive many retweets; it is not a signature of true spoiling of user interest by excessive friend exposure.  More specifically, because of the finite length of the dataset, some fraction of these will occur near the end of the collection period, and subsequent retweets by the user won't be observed, giving a false negative.  This effect is most pronounced when many friends have tweeted, because, if many friends have tweeted a URL, it is likely that a long time has elapsed between when the first and last friend tweeted, making it more likely to be cut off.  Additionally, people who have received more tweets are more likely to have more friends, putting them in a different population of user behavior, described below. This interpretation is additionally verified by the exposure response function, detailed in section~\ref{sec:trf}.

When we analyze the friend response function of  users based on how many friends they have, we see that user retweet probability is strongly dependent on the number of friends a user follows. Figure~\ref{fig:twitterFriendResponseBreakdown} shows the friend response functions for three populations of users, composed of slightly social users (who follow  1--10 friends), moderately social users (50--80 friends) and highly social users (more than 250 friends).
We see that the number of friends influences a user's sensitivity to incoming tweets.  On the face of it, the friend response functions in figure~\ref{fig:twitterFriendResponseBreakdown} suggest users with many friends may be jaded or disinterested, because highly connected users are less likely to retweet than poorly connected users, given an equal number of friend stimuli.  However, if instead we measure response to tweet \emph{exposures}, rather than the number of friends tweeting, we observe a pattern that suggests simpler behavior and rules out population-wide jadedness.

\subsubsection{Exposure response function}\label{sec:trf}
The \emph{exposure response function} gives the probability that a user will tweet a URL given the total number of times a URL appears in that user's Twitter queue.  It can be different from the friend response function, because Twitter allows users to tweet the same URL (or any message) multiple times, and each tweet will appear separately.
To find the total tweet exposure, we employ the direct sum, instead of an intersection,
\begin{displaymath}
V^{+,tw}(u,j) =\bigoplus_{\substack{i:t_i<t^*_{u,j} \\ u^\prime:u^\prime\in F^R_u}} C_{j}[u^\prime].
\end{displaymath}
Similarly, the set of tweets received by a user who did not tweet is
\begin{displaymath}
V^{-,tw}(u,j) =\bigoplus_{\substack{i:t_i<\infty \\ u^\prime:u^\prime\in F^R_u}} C_{j}[u^\prime],
\end{displaymath}
given $u\in \mathcal{W}_j$. The exposure response function is therefore given by
\begin{align}\label{eq:tweetresponsefunction}
P_{tw}\left(n|\chi\right) &= \\\nonumber
 &\sum_{u,j}\mathbbm{1}_{u\in\chi}\mathbbm{1}_{|V^{+,tw}(u,j)| = n}\bigg/ \\\nonumber
 \sum_{u,j}\mathbbm{1}_{u\in\chi}&\mathbbm{1}_{|V^{+,tw}(u,j)| = n}+\mathbbm{1}_{u\in\chi}\mathbbm{1}_{u\in\mathcal{W}}\mathbbm{1}_{|V^{-,tw}(u,j)| = n}.
\end{align}

\begin{figure}[tb] 
   \centering
  	 \includegraphics[width=3.5in]{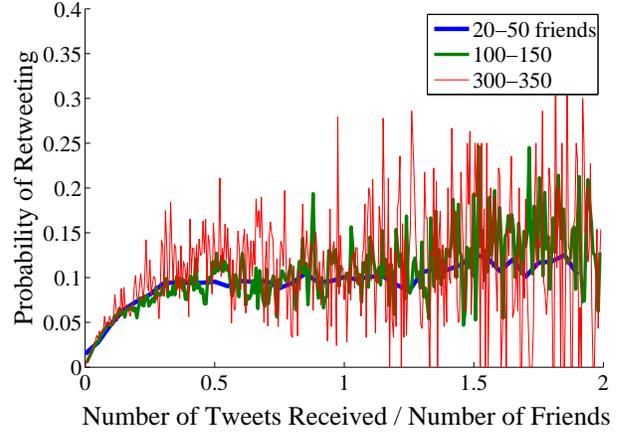}
   \caption{Relative exposure response  function of sub-populations of users.  Number of tweets normalized by the number of friends is a measure of the visibility of each URL  relative to all other incoming tweets. }
  \label{fig:tweetresponse}
 \end{figure}

Figure~\ref{fig:tweetresponse} shows the exposure response function for the  three populations. When reanalyzing the data based on the total number of received tweets, instead of number of tweeting friends, all exposure curves align (Users with many friends appear to be more active and more eager to retweet in general, explaining the slight excess retweet probability of high-friend individuals). This  alignment suggests a simple contagion mechanism in which a user's response is  proportional to the amount of URL-containing stimuli she receives relative to the total incoming stimuli, where the stimulus is simply seeing the URL in her the Twitter queue.  Because users' use Twitter intermittently, they only sample the incoming tweet queue. The more times the URL appears in a user's queue, the more likely she is to read and subsequently retweet it.



To isolate the effects  contributing to the observed retweeting behavior, we study how users respond to a single exposure to a URL.
This removes potentially confounding effects arising from multiple exposures,  such as memory or nonlinear perception.
We show that two mechanisms --- visibility and divided attention --- explain retweeting behavior. 

\begin{figure}[htb] 
   \centering
  	 \includegraphics[width=3.2in]{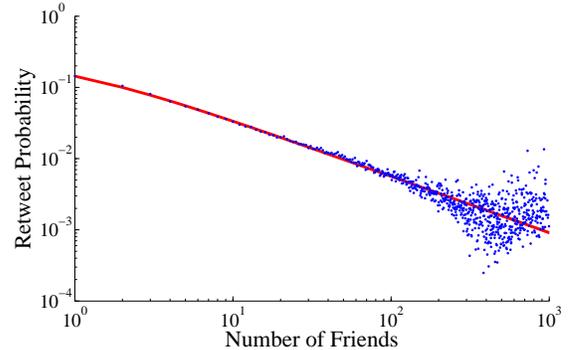}
   \caption{The probability of retweeting a URL given a single exposure drops off  monotonically as a function of the number of friends the user follows, denoted $n_f$.  The fit (red  line),  given by \eqref{eq:friendnormalization} in the text,  shows that the probability varies as approximately $n_f^{-1}$, which indicates that the user's attention is divided among incoming tweets.
   }
 \label{fig:friendnormalization}
   \end{figure}

\subsection{Effect of Divided Attention}
\label{sec:attention}
Above we showed that users' sensitivity to incoming tweets depends on  the number of friends they have. We characterize this empirically by measuring probability of retweeting a URL as a function of the number of friends the user has. This probability is computed for users who were exposed once and only once to the URL.  Figure~\ref{fig:friendnormalization} shows that this probability decays with the number of friends, denoted $n_f$. The solid line is the fitted function
\begin{equation}\label{eq:friendnormalization}
\mathcal{P}_{n_f} = 0.22\,{n_f^{0.21}}/{(n_f+0.53)}.
\end{equation}
 The factor of $n_f^{0.21}$ produces a small correction accounting for the higher Twitter activity in users with more friends.  The friend-dependent  probability of retweeting is dominated by the inverse of the number of friends.   Although in the present work we do not know all of the tweets a user receives,  total tweet arrival rate will be statistically proportional to $n_f$.

 Therefore, we attribute decrease in retweet probability with increasing $n_f$ to finite attention, the mechanism  controlling how people process incoming stimuli (including tweets) and decide which stimuli to act upon~\cite{Kahneman73}.   Because attention requires mental effort, and due the human brain's limited energy budget for executive functioning~\cite{Muraven:1998ta,Gailliot:2008cf}, the ability to process new stimuli is often rate-limited~\cite{blus:2008tn}, especially when considering the seemingly innumerable non-Twitter-related tasks simultaneously competing for attention.
Previous research has shown that attention limits the number of meaningful relationships people maintain online~\cite{Goncalves11}.
Our work shows that finite attention also affects short-term social contagion.
Users must divide their attention among  incoming tweets as they decide which tweets to propagate further.
Even if the user closely follows the tweets of specific friends, the user must still find  the influential tweets among the total accumulated tweets.
The more friends a user has, the more this problem is exacerbated, and the probability to respond to any individual friend's tweet is proportionally reduced.

\subsection{Effect of Decaying Visibility}
\label{sec:visibility}
Next we analyze the dynamic temporal response of users to a single exposure to a URL. We measure the \emph{time response function}, i.e., the probability of retweeting a URL
after a given delay, $P_t(t-t_0)$.
In analogy to the impulse response function in linear response theory~\cite{citeulike:5280499,van2007stochastic},  the time response function statistically describes how an average individual is stimulated to act following exposure to one tweet.
Analyzing this human response dynamic allows us to observe how users manage the inherently dynamic nature of the Twitter queue.
Although we do not know very much about any individual user, the large population size allows us to average out tweet-specific factors, revealing behavioral traits statistically common to the population as a whole.

\begin{figure}[tb] 
   \centering
  	 \includegraphics[width=3in]{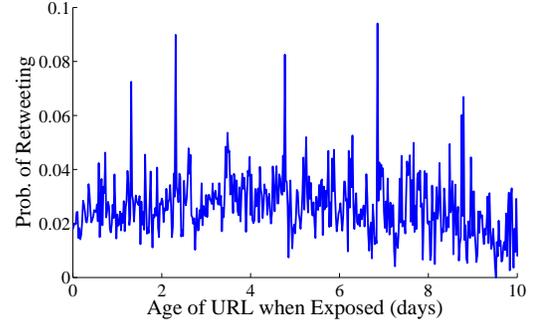}
   \caption{The probability of retweeting given a single exposure to the URL at the indicated age of the URL, in days, with 30 minute binning.  The age of the URL has little influence on the interestingness of the URL. We observe no decay of novelty, at least up to 10 days after appearance.}
   \label{fig:noveltydecay}
\end{figure}

The probability $P_t(t-t_0)$ is formally defined as $\langle \mathbbm{1}_{u_i,t} \mathbbm{1}_{u_i^\prime,t_0}\rangle_{u,t_0}$, where $\mathbbm{1}_{u_i^\prime,t_0}=1$ if user $u_i$ receives a tweet at time $t_0$ and 0 otherwise and given that $u_i$ follows $u_i^\prime$ and $t>t_0$.
To calculate the impulse response function, we first select all retweets that occur after receiving a URL only once.  A normalized histogram of this set of retweets gives the probability of retweeting after time $\Delta t=t-t_0$, given that a tweet arrived at time $t_0$ \emph{and} given that the user did retweet. We denote this conditioned probability distribution  as $\mathfrak{I}_\chi(\Delta t)$, where $\chi$ is the condition satisfied by the population of users under consideration. This gives the temporal envelope of the response probability, with unit normalization.  However, there is less than 100\% probability that a user will reply to a given tweet.  The proper normalization is given by calculating the probability that a user with characteristic $\chi$ eventually retweets, given that the user received only one tweet containing that URL,
\begin{align}\label{eq:timenormalization}
\mathcal{P}_{n_f} &= \int_0^\infty P_t(\Delta t|\chi)\,d\Delta t  = \\\nonumber
& \frac{\sum_{u,j}\mathbbm{1}_{u\in\chi} \mathbbm{1}_{|V^{+,tw}(u,j)| = 1}}{\sum_{u,j}\mathbbm{1}_{u\in\chi}\mathbbm{1}_{|V^{+,tw}(u,j)| = 1}+\mathbbm{1}_{u\in\chi}\mathbbm{1}_{u\in\mathcal{W}}\mathbbm{1}_{|V^{-,tw}(u,j)| = 1}}.
\end{align}
We calculated this normalization as a function of number of friends, shown in figure~\ref{fig:friendnormalization} and \eqref{eq:friendnormalization}.  To confirm that $\mathcal{P}_{n_f}$ is an equilibrium property, we calculated the normalization as a function of time evolved since the very first tweet in the time-series, i.e., the probability of retweeting given the age of the URL, $\tau$:
\begin{align}\label{eq:timenormalizationNovelty}
\int_0^\infty &P_t(\Delta t|\tau)\,d\Delta t  = \\ \nonumber
&\frac{\sum_{u,j} \mathbbm{1}_{|V^{+,tw}(u,j|\tau)| = 1}}{\sum_{u,j}\mathbbm{1}_{|V^{+,tw}(u,j|\tau)| = 1}+\mathbbm{1}_{u\in\mathcal{W}}\mathbbm{1}_{|V^{-,tw}(u,j|\tau)| = 1}},
\end{align}
\noindent where $\tau$ is the time since the first tweet containing the URL. $|V^{+,tw}(u,j|\tau)|$ is the number of tweets received by a user $u$ containing URL$_j$,
given that the last tweet received by $u$ containing URL$_j$ was at $\tau$ and $u$ did retweet URL$_j$. Equivalently, $|V^{-,tw}(u,j|\tau)|$ is the number of tweets received by a user $u$ containing URL$_j$,
given that the last tweet received was at $\tau$ but $u$ did not retweet URL$_j$.   As shown in figure~\ref{fig:noveltydecay}, the absolute age of the URL upon discovery by a user has little effect on the URL's retweet probability.

\begin{figure}[tbp] 
   \centering
   \subfigure[Blue: All users. Green: Low Connectivity Users. Red: High Connectivity Users. As a guide for the eye, the dashed line is $\propto\Delta t^{-1.15}$.]{
  	 \includegraphics[width=3.25in]{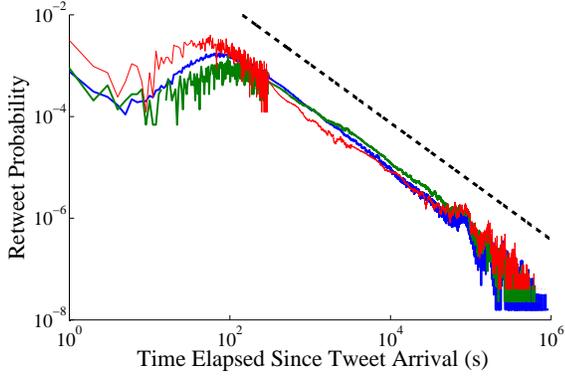}
 	  \label{fig:timeresponseall}
   }
   \subfigure[Dashed line is $\propto\Delta t^{-0.5}$.]{
  	 \includegraphics[width=3.25in]{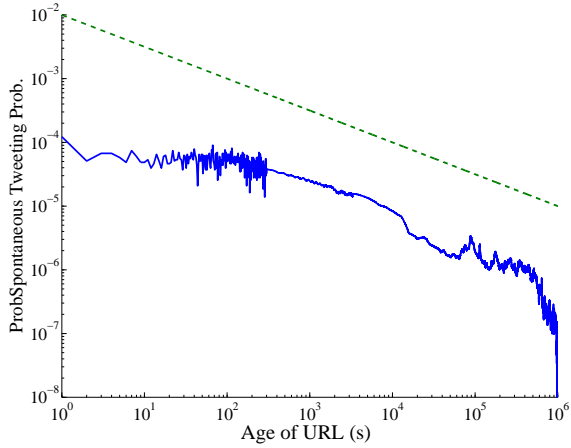}
 	  \label{fig:spontaneoustimeresponse}
   }
  \caption{(a) The time response function for users who retweet the URL after a single friend tweeted it and (b) the time dependence for users who spontaneously tweet the URL without  having any friend tweet it to them first.  All probabilities have unit normalization and are smoothed as follows: Between 0 and 300 seconds is raw data; between 301 and 33,000 seconds is a 300 sec. moving average; and from 33,001 onward is a 3000 sec. moving average.
}
   \label{fig:timeresponse}
 \end{figure}

Figure~\ref{fig:timeresponseall} shows the  time response function  $\mathfrak{I}_{\chi}(\Delta t)$\remove{, which is the probability of retweeting at a time relative to a user's first and only exposure to  a URL  given that the user did retweet it,} for $\chi = \{ \text{all users}, n_f < 10, \text{ or } n_f > 250\}$.   On average, users retweet the URL only minutes after a friend has tweeted it.  There is a broad peak around 2 minutes, which likely corresponds to users retweeting the URL after they read the referent webpage.
 After the characteristic read time, the time response function drops off roughly as $\Delta t^{-1.15}$, shown for comparison as a dashed line, and consistent with other observed power-law tails~\cite{Macskassy:2011uk,Eckmann:2004wm}.
 The time response function also shows modest revivals corresponding to circadian rhythms of user activity~\cite{citeulike:5839088}.

Figure~\ref{fig:spontaneoustimeresponse} shows the time-dependent response of users who did not receive the URL from a friend but tweeted it spontaneously (for example, after finding it at an external site.).   In calculating this spontaneous tweet probability, we take $t_0$ as the time of the first appearance of the URL in our sample.  Although information can spread from user to user outside of Twitter, the difference in power-law tails between spontaneous tweeting and the time response function  indicates that we are indeed observing retweeting  as claimed -- not chance independent discovery of the same URL elsewhere on the internet by two friends~\cite{Crane:2008hm}.
The probability to spontaneously tweet a URL remains relatively uniform over a period of about an hour, and then it decays with a power-law of approximately $t^{-0.5}$.  It does not show the peak around 2 minutes seen in figure~\ref{fig:timeresponseall}, reinforcing the interpretation that the time response function shows users often consume the URL before retweeting it.  Hence, the time response function contains the average of all the various human dynamics contributing to the retweet decision, including  how frequently users check Twitter, how they browse their Twitter queue, and how they consume a tweet's underlying content.


\remove{
\begin{figure}[tbp] 
   \centering
   \subfigure[]{
   \includegraphics[width=2.5in]{./Figures/timeResponseShortTime.pdf}
   \label{fig:timeresponseshort}
   }
   \subfigure[]{
  	 \includegraphics[width=2.5in]{./Figures/timeResponseLongTime.pdf}
 	  \label{fig:timeresponselong}
   }
   \caption{The time-dependent response function for users with three different amounts of friends. The dashed line is $\Delta t^{-1}$. (a) shows the response in the first 100 seconds, (b) shows the response at long time, with 1 minute smoothing.}
   \label{fig:timeresponse}
 \end{figure}

}

\begin{figure}[tbp] 
   \centering
  	 \includegraphics[width=3.1in]{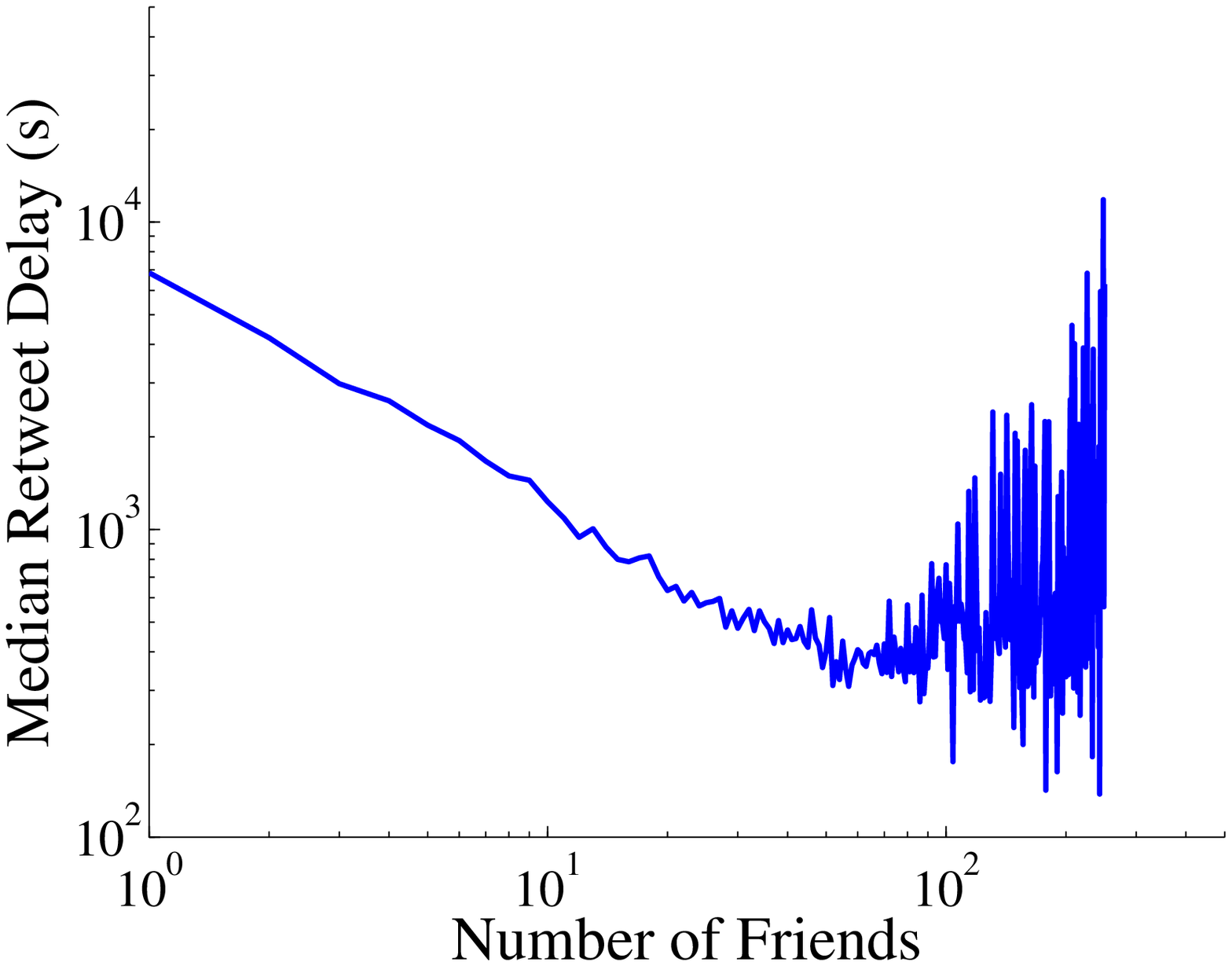}
  \caption{Median response time as a function of the number of friends.  Given that a user retweets a URL received once, competition among incoming tweets forces users with more friends to address incoming tweets more rapidly -- or risk having them  buried in their queue. }
 \label{fig:mediantimeresponse}
 \end{figure}

We interpret the time response function as decay of `\emph{visibility}' of a URL.  Figure~\ref{fig:timeresponseall} shows that the URL is most likely to be retweeted when it is most `visible' to the user, soon after a friend has tweeted it.
As soon as a putative URL arrives in a tweet, new tweets push the putative URL  down the user's Twitter queue, causing its saliency to decay.  Once the URL is no longer readily visible, the user must expend greater effort to find it, reducing the likelihood that it will be retweeted at all.
 Moreover, this effect is more dramatic when a user has more friends. Figure~\ref{fig:mediantimeresponse} reports the median response time of users to retweet a URL as a function of the number of friends they follow.
  Given that a user retweets a URL, a highly connected user with 50 friends does so more quickly than a poorly connected user with few friends.
 We know from the friend-dependent normalization, $\mathcal{P}_{n_f}$ that increased activity by highly connected users is insufficient to explain the linear decrease in median retweet delay.
 However, an explanation is readily provided by considering  decay of visibility: the temporal window of opportunity for easily finding a single tweet  becomes narrower as the number of friends grows.
 So, given that the user did retweet, they must have done it earlier if they had more friends, otherwise they are far less likely to have found the the tweet to begin with.
   This interpretation is further strengthened by figure~\ref{fig:timeresponseall}, where we see that highly connected users (250+ friends) are more likely to respond at short times and less likely to respond at long times, relative to low-connectivity users ($<10$ friends), yet the power-law exponent in the tails are nearly identical, indicating that the same underlying behavioral strategies are at play for all classes of users~\cite{citeulike:70828}.

\subsection{URL `Interestingness'}
\label{sec:url}
Equations \eqref{eq:timenormalization} and \eqref{eq:timenormalizationNovelty} deal with calculations averaged over all URLs.  However, some URLs are more compelling  than others.  We define the interestingness factor, $\mathcal{I}_j$, as the  scaling which determines the probability that a random user will retweet URL$_j$ given a single exposure.\footnote{We do \emph{not} propose that the content of a URL is completely characterized by a single real number -- only that the given quantity, which averages over all URL-specific factors, is one way to complete the description of the population-wide statistics, making interestingness similar to transmissibility.} That is,
\begin{equation}\label{eq:responsedef}
P_{j,t}(t-t_0,n_f) = \mathcal{I}_j \mathcal{P}_{n_f}\mathfrak{I}_{n_f}(t-t_0),
\end{equation}
\noindent where $n_f$ is the number of friends and $\mathcal{P}_{n_f}$ is the normalization for the class of user with $n_f$ friends, shown in figure~\ref{fig:friendnormalization}.  As shown in figure~\ref{fig:timeresponseall}, the precise time dynamics of retweeting probability depends on $n_f$, hence the expression $\mathfrak{I}_{n_f}.$ For URL$_j$, we consider only those users who were exposed to  URL$_j$ from precisely one tweet, denoted by the set $\mathfrak{N}_{1,j}$.  The expected number of retweets by  all users  exposed once  to URL$_j$ is $\langle {N}_{1,j}\rangle \equiv \sum_{i\in\mathfrak{N}_{1,j}} \mathcal{I}_j\mathcal{P}_{n_f,i},$ where $\mathcal{P}_{n_f,i}$ is the normalization for $u_i$, given in \eqref{eq:friendnormalization}. Therefore, the estimator for $\mathcal{I}_j$ is $\langle \mathcal{I}_j \rangle = N_{1,j}/\sum_{i\in\mathfrak{N}_{1,j}} \mathcal{P}_{n_f,i},$ where $N_{1,j}$ is the observed number of retweeting users exposed once to URL$_j$.

\begin{figure}[tbh] 
   \centering
    \subfigure[Solid line is log-normal fit with $\mu=-0.13$ and $\sigma=0.92$.]{
 	 \includegraphics[width=3.25in]{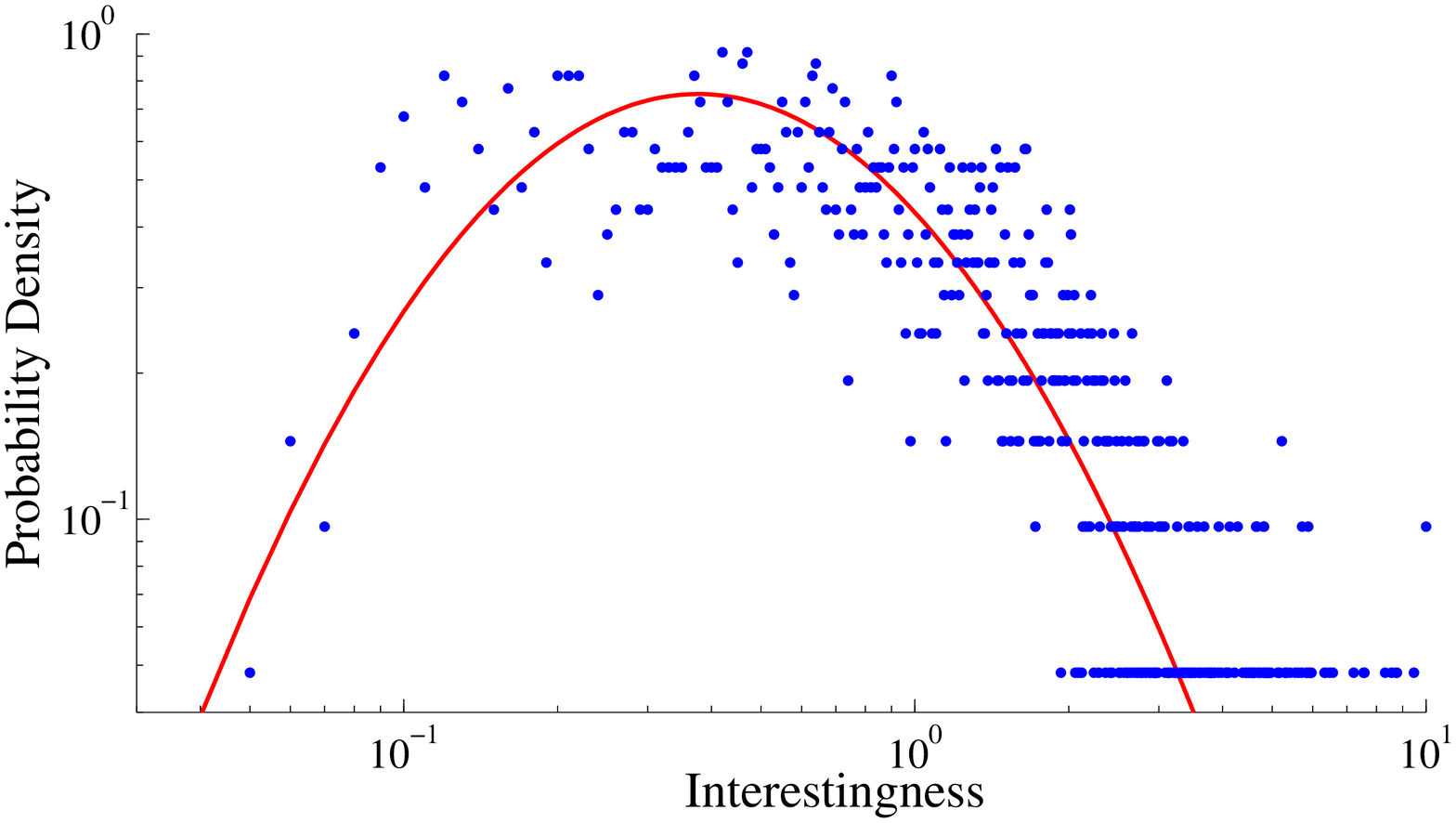}
    \label{fig:interestingnessA}
    }
    \subfigure[Red is most likely, and Blue is the least likely.]{
 	 \includegraphics[width=3.25in]{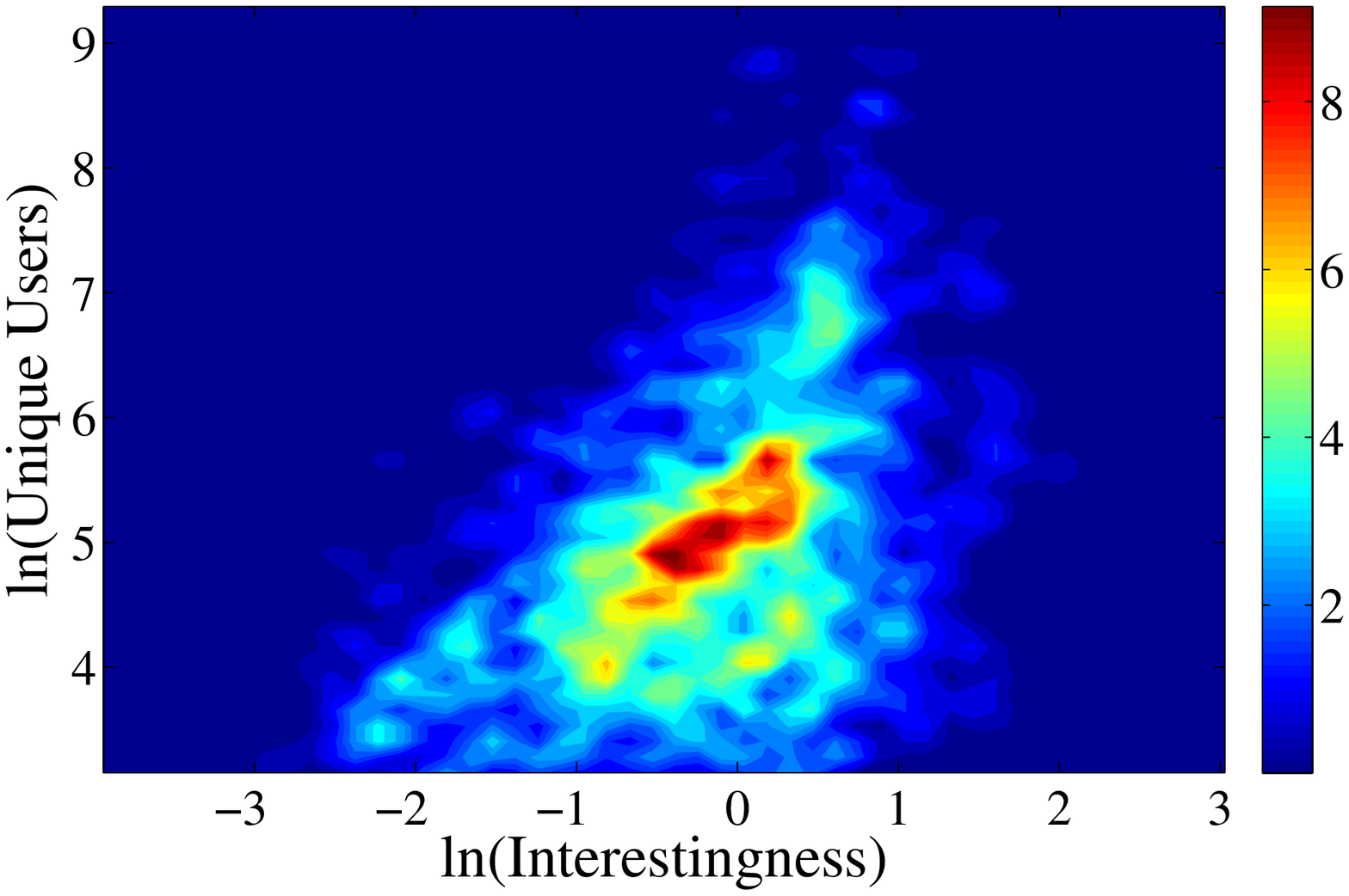}
    \label{fig:interestingnessB}
    }
    \caption{URL `interestingness', i.e., the scaling factor determining the probability that a random user will retweet a URL. (a) Distribution of estimated  interestingness values of URLs in the data set. The line represents a log-normal fit. (b) A density plot of URL interestingness vs. the number of users tweeting the URL. }
  \label{fig:interestingness}
 \end{figure}

Figure~\ref{fig:interestingnessA} shows the distribution of estimated interestingness values of URLs in our data sample. The solid curve gives a log-normal fit to the data.  A log-normal distribution of  interestingness was also observed for news stories on Digg after controlling for their visibility~\cite{Lerman10www}. This suggests that similar processes may be at work on both sites in selecting (or creating) interesting content.

One question we can ask is whether more interesting or `contagious' URLs reach a wider audience, measured by the number of unique users who retweet them.  Figure~\ref{fig:interestingnessB} plots the density of observed occurrences of the dyad (URL interestingness, number of unique users who tweeted URL). Though there is some positive correlation in the two quantities, the dependence is much smaller than we expected; the number of people who can be reached by URLs of similar interestingness can vary by five orders of magnitude! This is not inconsistent with  measures of interestingness in other contexts, such as website traffic vs. incoming links~\cite{Fortunato06pnas}.  Not all highly compelling information propagates far in the network, and conversely, sometimes even uninteresting information can spread very far.  Understanding the specific mechanics of social contagion is necessary for identifying the conditions for interesting content to go viral. 

%% file: related-work-final.tex
\section{Related Work}\label{sec:related work}

Recent research suggests that  social contagion is a complex process that depends on the type of information being transmitted~\cite{Romero:2011um}, its novelty~\cite{Wu07}, influence of the initial spreader~\cite{Bakshy11} and activity levels of subsequent spreaders~\cite{citeulike:3680249}, how people respond to repeated exposure to information by their network neighbors~\cite{Romero:2011um,Versteeg11,Centola10,Mason12}, as well as network structure~\cite{Centola10,GonzalezBailon:2011kz,Mason12}.  We show that most of the retweeting probability can be simply explained in terms of two factors: visibility and attention, which has recently emerged as a new factor to consider in explaining mechanics of social contagion.  Previous work on attention has focused on \emph{collective} attention, the attention a meme received from the entire community~\cite{Lehmann11,Moussaid09,Wu07,citeulike:9266822,Ratkiewicz10}.  These works describe how the community as a whole shifts its attention from one meme to another, albeit acknowledging the finite nature of individual attention.  This line of analysis has begun to take much more careful consideration of how divided attention limits the information processing capability of individuals.  For example, Goncalves et al.~\cite{Goncalves11} showed that the maximum number of other users that Twitter users converse with is bounded by 100-200, which is close to the cognitive limit known as ``Dunbar's number.'' Also, Weng et al. proposed a heuristic model of finite memory (in both capacity and time) to account for retweeting of hashtags and hashtag lifetime~\cite{Weng:2012dd}.  The present work does not rely on a heuristic of finite memory, as we draw upon the well documented principle of least effort in the context of the biological energy demands of balancing Twitter use with all of the other tasks we must address in our daily lives.  We conclude that well-connected individuals have greater difficulty in spreading information due to difficult in sampling the denser incoming information stream, something only indirectly accounted for in the finite memory model.  Additionally, our analysis of visibility decaying due to spatial competition in the user interface provides an experimentally verifiable hypothesis, unlike other pictures of dynamic meme competition, such as the previously mentioned finite memory model or the ``decay of novelty" hypothesis.

The dynamic deterioration of retweeting probability is sometimes explained as ``decay of novelty," which refers to the model that a URL's inherent interestingness decreases as it ages~\cite{Wu07}. There are some problems with this interpretation.  First, while novelty may apply to sports or weather, which appeals largely due to its timeliness, the present work's observed dynamic decrease in retweeting probability  applies to all content.  Second, retweet probability depends only on the time evolved since a user's friend tweeted it (via the \emph{time response function}), not the  absolute age of the URL.  Figure~\ref{fig:noveltydecay} shows the probability of retweeting a URL after a single exposure vs. the age of the URL, i.e., the time since its first appearance on Twitter, and this probability remains nearly constant over the time-scale of inquiry.
Hence, for most URLs, the absolute age of the URL does not correlate with any decay of novelty.
Finally, novelty as a model-specific parameter is an inherent property of the information represented by the URL~\cite{Asur:2011tc}, while visibility is a property of the user interface.   When old information is given high visibility, it can have the same virality as novel information.  An interesting example occurred in 2008 when a mistake by the Google News algorithm led to a 2002 story about United Airlines filing for bankruptcy protection to be featured as its top news. United Airlines stock price plummeted as people reacted to the erroneously publicized  old news~\cite{Carvalho:2011wo}.  

%% file: discussion-final.tex
\section{Conclusion}~\label{sec:discussion}
The present work analyzed human response dynamics to understand observed Twitter use.
By considering URL transmission in tweets, we infer that the first-in-first-out queue structure of the Twitter interface is responsible for the characteristic temporal decay of visibility and complimentary enhancement of visibility immediately after receiving a URL-containing tweet. 
Visibility is quantitatively manifested via  the revival of the probability of tweeting each time a new tweet  arrives, according to~\eqref{eq:responsedef}, because it takes the user less effort to discover a tweet at the top of the queue.  Visibility not only impacts retweet selection, but it helps explain other observations,  such as click position bias~\cite{Fortunato06pnas,Craswell08}.
We also presented evidence that users divide their attention among friends, which limits their ability to respond to an arbitrary tweet.
Attention is manifested by normalizing  the overall probability of tweeting by number of friends, as shown in figure~\ref{fig:friendnormalization}.
The combination of dynamic visibility and limited attention provides a foundation for explaining social dynamics which is more consistent with the data than the decay of novelty or collective attention models. 

Our study points to the fundamental role limited attention and visibility play in the spread of information in online social networks.  It also reveals the inherently dynamic nature of social contagion -- ideas compete in time and space for user adoption.
The probability of a meme being spread depends on the product of visibility and attention. The precise mechanisms of this interplay will depend on the underlying contagion model, but the present work demonstrates how the principle of least effort clearly plays a key role in determining behavior on Twitter.  Even when interesting content  has a very high probably of being retweeted,  rapid visibility decay and finite attention causes information to be effectively lost to the user, because of the excess effort required to act on the tweet compared with younger -- more visible -- information, explaining why most content fails to go viral.  

\remove{ From a theoretical perspective, the simplest and most widely studied model of social contagion is the independent cascade model, which is often used to model the spread of a disease from infected people to their friends~\cite{newman02,citeulike:5087132}. In this model, each exposure of a healthy person by an infected friend leads to an independent chance of transmitting the disease and becoming infected.  The  probability of infection increases monotonically with the number of infected friends~\cite{Versteeg11,Castellano:2009cea}.  The present results suggest a modified independent cascade model would be better suited for social contagion.  In the modified model, infection rate is inversely proportional to in-degree (attention) and infection transmission follows   non-Poisson temporal dynamics (visibility), such as  in figure~\ref{fig:timeresponseall}.}

Visibility and attention are highly interlinked: the fact that we have finite attention makes visibility much more important in arousing a response~\cite{Melcher:2011eq}.
However, although attention is inherently a property of an individual user, social media providers can manipulate it through the design choices they make in the user interface. For example, the online social news aggregator Digg streamed friends' recommendations as a chronological queue, ordered by time of first recommendation. The first time a news story is recommended to users, it appears at the top of their queues. Recommendations for other news stories cause the visibility  of an older story to decrease, as the story moves down their queues. Subsequent recommendations for the same story do not change its position within the queue, so that by the time several more friends have recommended the story, it has moved so far down the queue as to be virtually invisible to the user. This potentially explains the failure of Digg users to respond to multiple recommendations, which was shown to drastically limit how far news stories spread on Digg~\cite{Versteeg11}. Contrary to Digg, Twitter orders posts by their latest activity, with the most recently retweeted URL at the top of the queue. This makes it  easier for users to respond to popular recommendations and changes the nature of contagion, though its gross properties remain similar to Digg~\cite{Lerman10icwsm}. Although user attention is fundamentally limited, interface design can manipulate visibility to maximize the quality of user choices and social interactions.  User interfaces could also directly alter the saliency of items of interest, by altering their color and size or by adding distinctive badges, to draw attention to social signals indicating potentially interesting information.


